\newcommand{\SY}{\scriptstyle}

\documentstyle[preprint,tighten,floats,pre,aps,epsf]{revtex}

\begin{document}
% \preprint{\fbox{\parbox{140mm}{First line,\\second line.}}}
%\preprint{\fbox{\parbox{90mm}{\footnotesize Corresponding author:\\
% \hspace*{2ex} Prof.~Dr.~D.~W.~Heermann\\
% \hspace*{2ex} Institut f\"ur Theoretische Physik, Universit\"at Heidelberg\\
% \hspace*{2ex} Philosophenweg 19, D--69120 Heidelberg, Germany\\
% \hspace*{2ex} Tel: +49--6221--54--9448, Fax: +49--6221--54--9331\\
% \hspace*{2ex} E--mail: {\tt heermann@tphys.uni-heidelberg.de}\\
% \hspace*{2ex} WWW: {\tt http://wwwcp.tphys.uni-heidelberg.de/}}}}
\draft
\title{Modelling of amorphous polymer surfaces in computer simulation}
\author{Thorsten Hapke, Andreas Linke, Gerald P\"atzold,
  and Dieter W.~Heermann\thanks{Contact:
  {\tt heermann@tphys.uni-heidelberg.de},
  {\tt http://wwwcp.tphys.uni-heidelberg.de/}}}
\address{Institut f\"ur Theoretische Physik, Universit\"at Heidelberg,
  Philosophenweg 19, D--69120 Heidelberg, Germany
  and Interdisziplin\"ares Zentrum f\"ur wissenschaftliches Rechnen
  der Universit\"at Heidelberg}
\date{\today}
\maketitle
\begin{abstract}
We study surface effects in amorphous polymer systems by means of computer
simulation. In the framework of molecular dynamics, we present two different
methods to prepare such surfaces. {\em Free} surfaces are stabilized solely
by van--der--Waals interactions whereas {\em confined} surfaces emerge in the
presence of repelling plates. The two models are compared in various computer
simulations. For free surfaces, we analyze the migration of end--monomers to
the surface. The buildup of density and pressure profiles from zero to their
bulk values depends on the surface preparation method. In the case of confined
surfaces, we find density and pressure oszillations next to the repelling
plates. We investigate the influence of surfaces on the coordination number,
on the orientation of single bonds, and on polymer end--to--end vectors.
Furthermore, different statistical methods to determine location and width
of the surface region for systems of various chain lengths are discussed
and applied. We introduce a ``height function'' and show that this method
allows to determine average surface profiles only by scanning the
outermost layer of monomers.\\[2ex]
\end{abstract}
% 68.10.-m  Fluid surfaces and fluid-fluid interfaces
% 36.20.Ey  Macromolecules and polymer molecules: Conformation
%           (statistics and dynamics)
% 61.20.Ja  Computer simulation of liquid structure
% 61.41.+e  Structure of solids and liquids: Polymers, elastomers, and plastics
% 68.15.+e  Liquid thin films

%%% ---------------------------------------------------------------------------
\section*{Introduction}
%%% ---------------------------------------------------------------------------

The technological applicability of polymeric materials depends in a
number of important cases on the polymer's surface properties: the surface
roughness determines the friction coefficient, cracks emerging from the
surface influence the material's durability, and surface defects reduce the
transparency of polymeric panes. At the same time, the experimental
and manufacturing techniques advance to smaller and smaller
scales. Nowadays one can pursue mechanical effects on the nano--scale.
This opens new, challenging, and exciting roads for academic research
and promising prospects for industrial applications. Nano--mechanics
influences the design of computer chips and is fundamental for new
developments like miniaturized optical devices and nano--engines.
A related topic is the mechanics of very thin films~\cite{Bog95a}
with thicknesses in the nano--range. But again, it is not just the bulk
behaviour that decides about the nano--technological applicability of a
certain material but its surface properties.
The nano--scale processes that lead to macroscopically relevant
surface--related phenomena like friction, wear, hardness, and certain types
of material failure like surface degradation are scarcely understood from the
theoretical point of view. In some cases, phenomenological theories
exist~\cite{Spu82a} which attempt to systematize the various experimental
observations. A fundamental understanding, however, tentatively based on
fundamental nano--scale processes, is lacking. The reason is that on this
scale, the number of mutually interacting particles is very high, but the
system is still too coarse to apply continuum mechanics in a straightforward
way. Some simplifications can be made in the case of crystalline solids but
amorphous systems like polymer glasses still represent a major challenge.
Techniques from computational physics can help to deal with the many degrees
of freedom involved~\cite{Hee86a}.

In ongoing work, we intend to address several basic questions that
pertain to the understanding of nanoscopic amorphous systems in conjunction
with tribological and surface effects like friction, wear, and hardness.
Our investigation is mainly restricted to polymeric materials. Crystalline
solids have been studied in a beautiful series of pioneering papers by
Landman et al.~\cite{Lan92a}. In the present contribution, we concentrate
on the preparation of polymer surfaces in the framework of computer
simulation. We are furthermore concerned with the detailed characterization
of important surface properties. This is a necessary foundation to correctly
interpret our findings from a large number of dedicated polymer surface
indentation simulations~\cite{Lin96a}. These simulations provide a prototype
nano--scale experiment to probe surface hardness, friction, and damage
mechanisms. Related to these activities are computer studies on the
deformability of nano--scale polymeric films~\cite{Pae96a}.

%%% ---------------------------------------------------------------------------
\section*{Polymer and surface modelling}
%%% ---------------------------------------------------------------------------

For the polymer chains, we use a united atom model in conjunction
with Newtonian dynamics. In addition to harmonic chain forces which
keep the bond lengths next to the equilibrium value, we model the
fluctuation of bond angles, again by a quadratic potential. Between
monomers which do not participate in mutual bond length or bond angle
interactions, Lennard--Jones forces are acting, both to model an excluded
volume effect and to hold the polymer system together. Note that we neglect
any torsional potential in the present study. To be explicit,
the Hamiltonian of the model is of the general form
\begin{eqnarray}
  H &=& H_{\rm \SY bondlength} + H_{\rm \SY bondangle} + H_{\rm \SY LJ}
    \, , \nonumber \\
  H_{\rm \SY bondlength} &=& \sum_{\rm \SY bonds} \frac{k_b}{2}
    \left( l_{\rm \SY bond}
    - l_0 \right)^2 \, , \nonumber \\
  H_{\rm \SY bondangle} &=& \sum_{\rm \SY angles} \frac{k_\theta}{2} \left( 
    \cos \theta_{\rm \SY angle} - \cos \theta_0\right)^2 \, , \\
  H_{\rm \SY LJ} &=& \sum_{\rm \SY pairs \; of \atop \rm \SY monomers}
    4 \epsilon \left[ \left( \frac{\sigma}{r_{\rm \SY pair}} \right)^{12} -
    \left( \frac{\sigma}{r_{\rm \SY pair}} \right)^{6} \right] \, \nonumber .
\end{eqnarray}
The Lennard--Jones interaction is implemented with a cutoff of
$2.5 \, \sigma$ and appropriate potential and force shifts are
used to retain continuity.
We use molecular dynamics methods to compute the motion of the monomers.
Models of this kind are described at various places in the
literature~\cite{All87a}. We intend to capture some essential
features of polyethylene chains and appropriate model parameters are
compiled in Tab.~\ref{table-01} (see also Ref.~\cite{Pau95a}).

\begin{table}
  \caption{Parameters of the polyethylene model.}
  \label{table-01}
  \begin{tabular}{lll}
    Lennard--Jones energy, $\epsilon$ & $8.3027 \cdot 10^{-22}$ J
      & 5.18 meV \\
    Lennard--Jones length, $\sigma$ & 380 pm
      & 3.8 {\AA} \\
    monomer mass (${\rm CH}_2$ group), $m$ & $2.3248 \cdot 10^{-26}$ kg
      & 14 atomic units \\
  \tableline
    unit of temperature, $\epsilon / k_{\rm B}$ & 60.1357 K \\
    unit of mass density, $m / \sigma^3$ & 423.6687 ${\rm kg}/{\rm m}^3$
      & 0.4237 ${\rm g}/{\rm cm}^3$ \\
    unit of time, $\left(m \sigma^2 / \epsilon \right)^{1/2}$ & 2.0108 ps \\
    unit of velocity, $\left( \epsilon / m \right)^{1/2}$ & 188.9822 m/s \\
    unit of force, $\epsilon / \sigma$ & 2.1849 pN
      & 1.36 ${\rm meV} / {\rm {\AA}}$ \\
    unit of spring constant, $\epsilon/\sigma^2$ & $5.7498 \cdot 10^{-3}$ N/m
      & 0.36 ${\rm meV} / {\rm {\AA}}^2$ \\
    unit of pressure, $\epsilon / \sigma^3$ & 151.3103 bar
      & 0.09 ${\rm meV} / {\rm {\AA}}^3$ \\
  \tableline
    temperature, $T$ & 301 K and 361 K \\
    bond length, $l_0$ & 152 pm
      & 1.52 {\AA} \\
    bond angle, $\theta_0$ & 109.47${}^{\text{\,o}}$
      & $\cos \theta_0 = - 1/3$ \\
    spring constant (bond length), $k_b$ & $5.7498 \cdot 10^{1}$ N/m
      & 3.59 ${\rm eV} / {\rm {\AA}}^2$ \\
    bending constant (bond angle), $k_\theta$ & $8.3027 \cdot 10^{-19}$ J
      & 5.18 eV \\
    simulation time step, $\Delta t$ & 2.0108 fs
  \end{tabular}
\end{table}

\begin{figure}
\epsfxsize=11cm
  \begin{center}
    \begin{minipage}{\textwidth}
      \epsffile{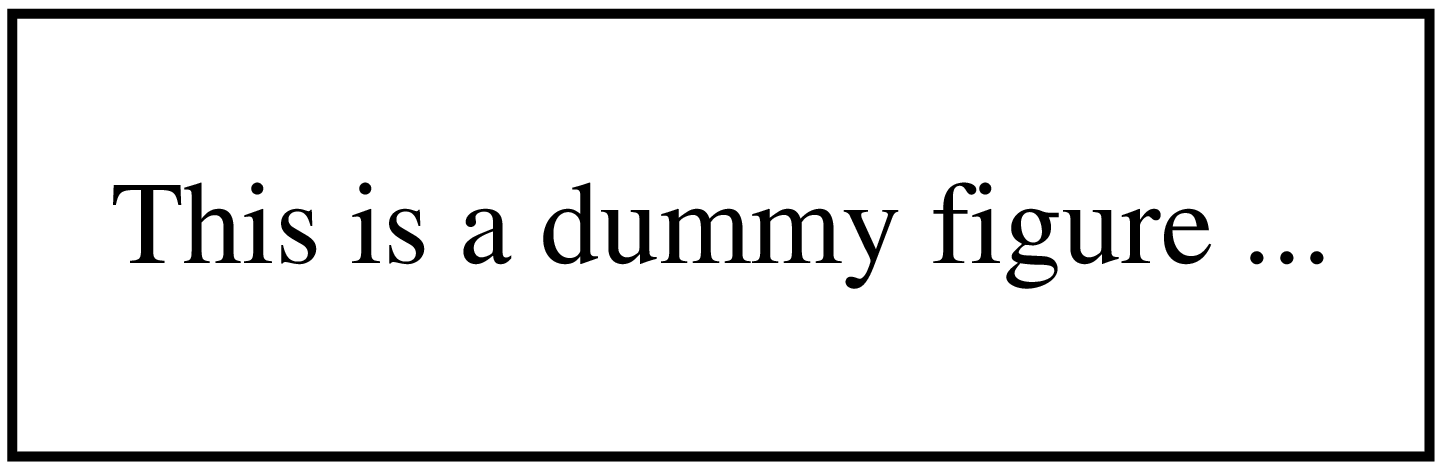}
      % \epsffile{confsurf.eps}
    \end{minipage}
  \end{center}
\caption{The {\em confined} polymer surface (at the top of the bulk).
  The horizontal line of the frame represents the repelling plate.
  The monomers are drawn with their Lennard--Jones radius $\sigma$.
  In the lateral directions, periodic boundary conditions apply.}
\label{fig:confined_model}
\end{figure}

\begin{figure}
\epsfxsize=10cm
  \begin{center}
    \begin{minipage}{\textwidth}
      \epsffile{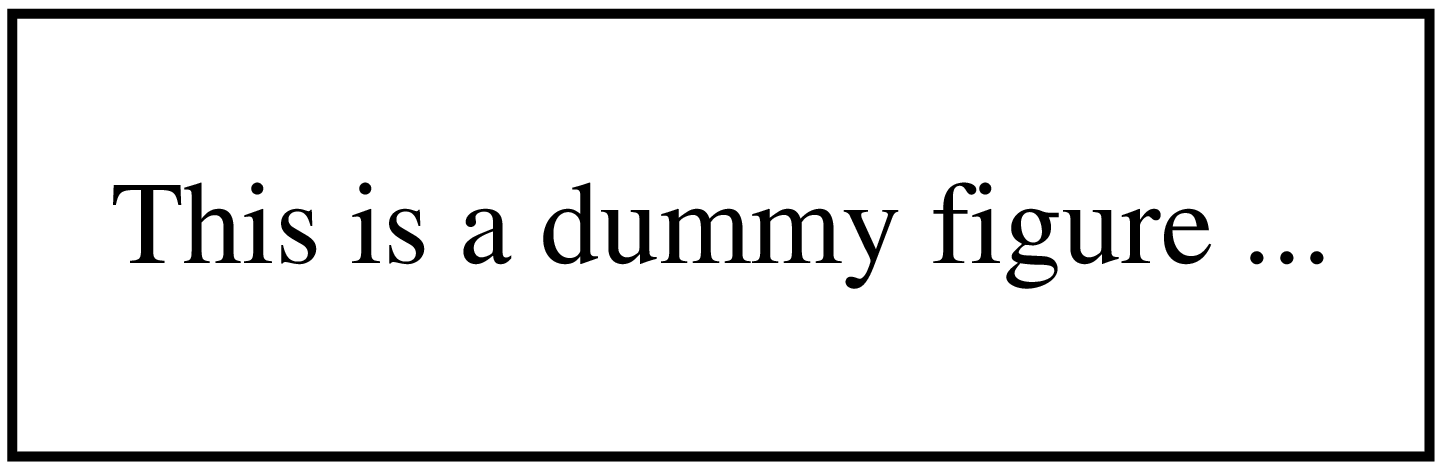}
      % \epsffile{freesurf.eps}
    \end{minipage}
  \end{center}
\caption{The {\em free} polymer surface (at the top of the bulk).
  The horizontal line of the frame is out of the interaction range.
  The monomers are drawn with their Lennard--Jones radius $\sigma$.
  In the lateral directions, periodic boundary conditions apply.}
\label{fig:free_model}
\end{figure}

We employ two alternative methods to prepare amorphous polymer surfaces. The
{\em confined polymer surface}, Fig.~\ref{fig:confined_model}, is realized
by plates that repell the monomeric units and thereby restrict the polymer
to a certain region in space. This gives rise to a surface model which is
mainly characterized by its smoothness. In the simulations, we set up the
united atom model with periodic boundary conditions in the $x$-- and
$y$--direction. In the $z$--direction, we put two repelling plates at
a distance chosen such that the resulting bulk density of the polymer system
settles to the desired value. The plates interact with the polymers through 
the repelling ($r^{-12}$) part of the Lennard--Jones potential.
The other alternative, Fig.~\ref{fig:free_model}, is the {\em free polymer
surface}. Here the polymer bulk is held together solely by the overall
attracting intermolecular van--der--Waals interactions. Individual chains
may evaporate or condensate. The emerging surfaces are rougher than in the
confined case. We suggest that the confined model applies to the interface
between the polymer and a comparatively rigid material like a metal. Such
surfaces may also be formed on explosion pressed materials. On the other hand,
the free surface is capable to model the interface between the polymer and a
(dilute) medium where any mutual interactions can be neglected (in effect, it
is the interface to a vacuum).

\begin{figure}
\epsfxsize=14cm
  \begin{center}
    \begin{minipage}{\textwidth}
      \epsffile{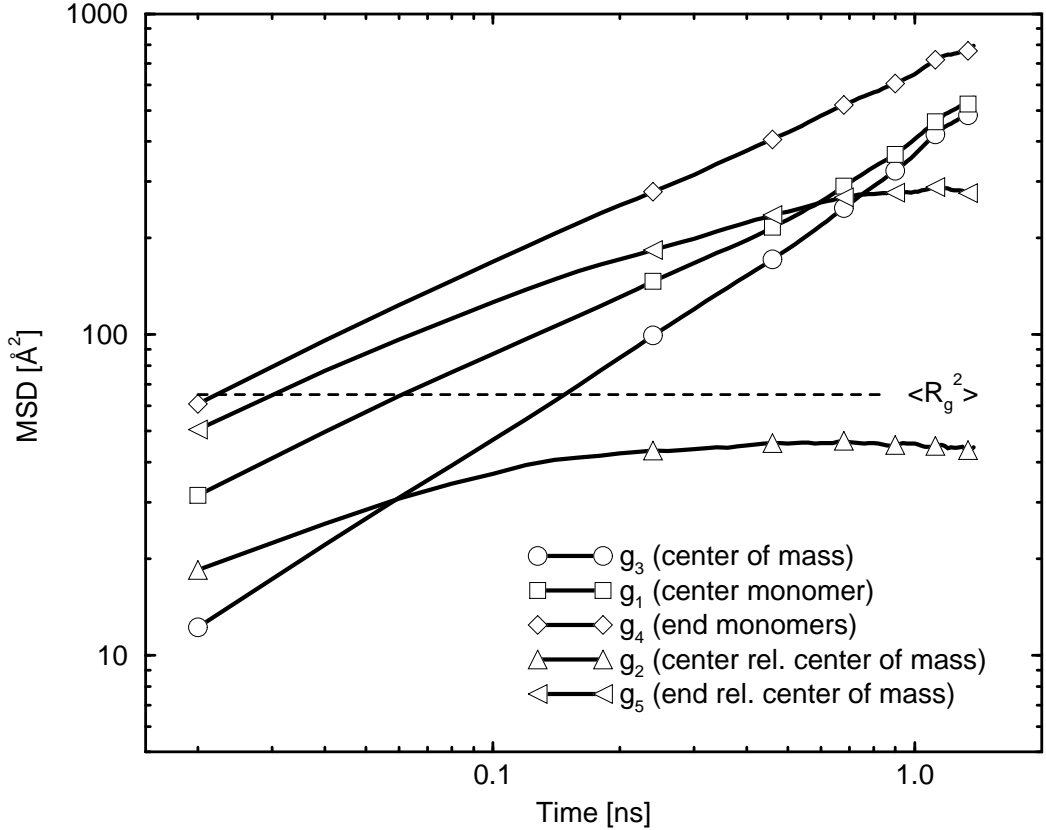}
    \end{minipage}
  \end{center}
\caption{Various measures of the mean square displacement (MSD).
  In particular, $g_1$ is the MSD of the center monomers, $g_2$ is
  the same but relative to each chain's center of mass, $g_3$ is the
  MSD of the chains' center of mass, $g_4$ is the MSD of the end
  monomers, and $g_5$ is the same but again relative to each chain's
  center of mass. The chain length is $N=40$ and only every tenth sample
  point is marked by a symbol.}
\label{fig:polymer_displacement}
\end{figure}

\begin{figure}
\epsfxsize=14cm
  \begin{center}
    \begin{minipage}{\textwidth}
      \epsffile{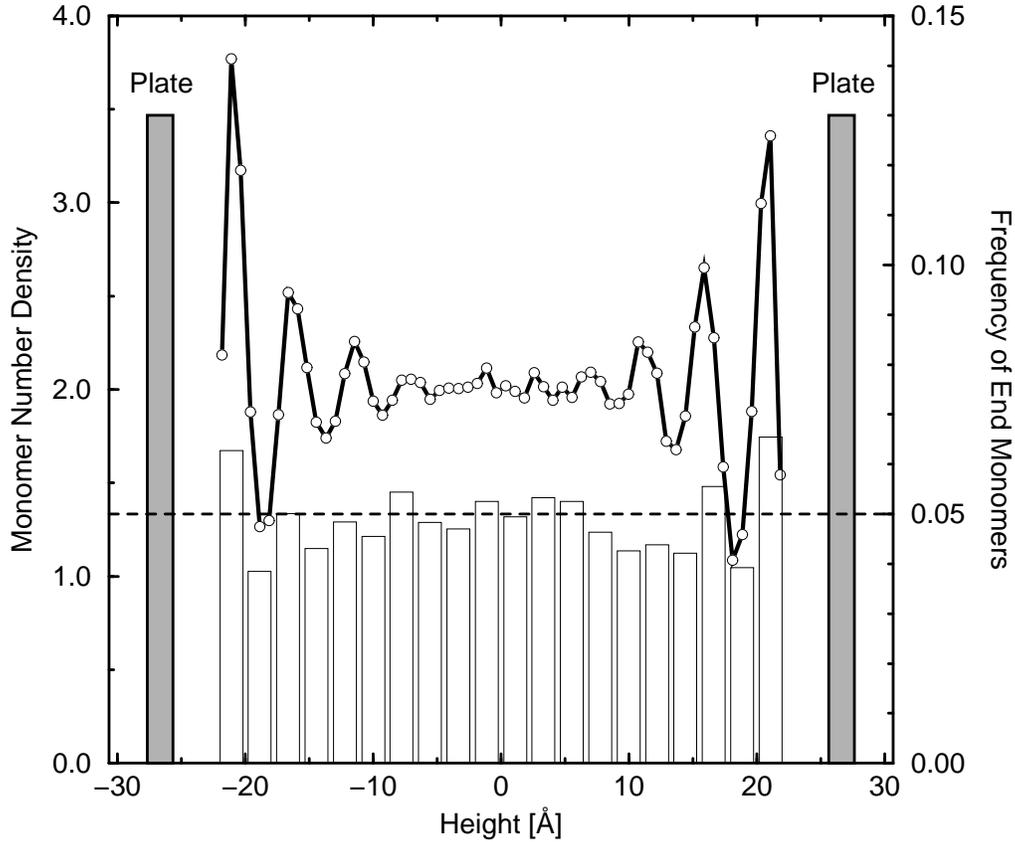}
    \end{minipage}
  \end{center}
\caption{Distribution of density and end--monomers for the
  {\em confined} surface. Shown is the number density of monomers
  (solid curve, left axis) and the relative frequency of chain--end
  monomers (bar chart, right axis) in slices parallel to the surface.
  The polymer system is monodisperse and contains 205 chains of $N=40$
  monomers each. The horizontal dashed line indicates the mean value
  for chain--ends: $\text{2 ends}/\text{40 monomers} = 0.05$. Also shown
  is the location of the confining plates at both sides.}
\label{fig:confined_model_density_dist}
\end{figure}

All simulations to probe surface properties are perfomed at reduced
temperatures of  $T^* = T / (\epsilon / k_{\rm B}) = 5.0$ (confined
system) or $6.0$ (free system).
These temperatures correspond to $301$ or $361$ Kelvin which is above the
melting temperature of real polyethylene. We stress that our simulations
are not yet supposed to fully mimic real polyethylene (it has been
demonstrated, however, that much can be achieved by carefully tuning the
parameters of united atom models~\cite{Pau95a}). The chain lengths
are still too short and we treat chain--end monomers like mid--chain monomers.
We also neglect the torsional potential (the rotation about C--C axes is not
restricted). The high temperature helps to ensure that in feasible simulation
times, the chains travel several radii of gyration in order to guarantee
correct thermodynamic properties.
To check this, various mean square displacement measures as defined in
Ref.~\cite{Pau91a} are shown in Fig.~\ref{fig:polymer_displacement}.
Indication that the chain in its entirety has moved through the matrix
comes from the relative displacements of the inner monomers relative to
the center of mass motion ($g_2$) and the outer monomers relative to the
center of mass motion ($g_5$). The cross--over points define charateristic
time--scales of the system. The presence of the cross--over points in
Fig.~\ref{fig:polymer_displacement} indicates that the simulation time
scale is sufficiently long compared to the short--order relaxation times.
Note that the data for all the figures to follow represent averages over
a great number of system configurations which have been sampled from
simulations which extend over a span of time comparable to that in
Fig.~\ref{fig:polymer_displacement}. For each chain length ($N\!=\!20$, $40$,
$60$, and $80$), the simulation follows 8200~monomers over 1.4~nanoseconds.
% Every 0.02~nanoseconds, a cofiguration was sampled, resulting in a total of
% 70~files to average over.
If not otherwise indicated, the results given are for the chain
length $N\!=\!40$.

%%% ---------------------------------------------------------------------------
\section*{Confined polymer surfaces}
%%% ---------------------------------------------------------------------------

In the case of the polymer system with confined surfaces, the plates are
rapidly brought from infinity to a distance such that the monomer number
density in the resulting box is $\rho_{\text{bulk}}=2.0$. The shock results
in a distribution of the density in the $z$--direction (system height) as
shown in Fig.~\ref{fig:confined_model_density_dist}. These typical periodic
changes of the density distribution close to the surface~\cite{Bra94a}
are due to the external pressure produced by the plates, which leads to
a local partial crystallization. The same behaviour has been found in
simulations analyzing surface phenomena~\cite{Rou96a} and studied in the
context of one long chain between two parallel plates~\cite{Oph85a}. The
chains experience strong packing constraints as well as a loss in entropy
that must show up also in the surface properties that can be observed during
indentation.

\begin{figure}
\epsfxsize=14cm
  \begin{center}
    \begin{minipage}{\textwidth}
      \epsffile{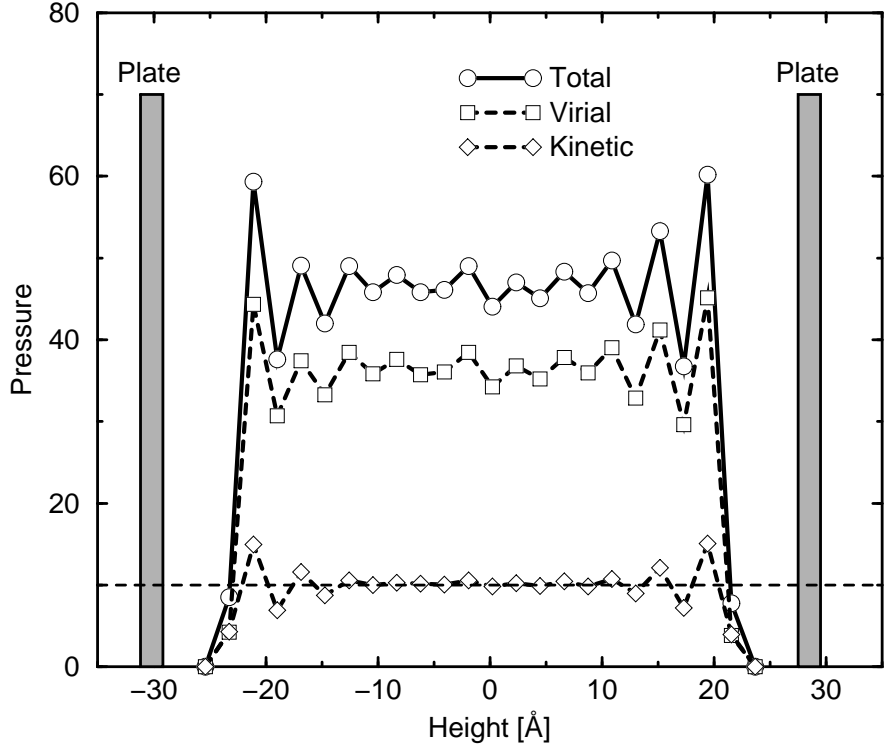}
    \end{minipage}
  \end{center}
\caption{The pressure distribution vs.~the height direction for the
  {\em confined} surface model (pressure in reduced units). The total
   pressure is the sum of the virial and the kinetic contribution. Also
   shown is the location of the confining plates at both sides. The dashed
   horizontal line indicates the expected value for the kinetic
   (ideal) part: $p^{*}_{\text{ideal}} = \rho^* T^* \approx
   2.0 \cdot 5.0 = 10.0$.}
\label{fig:confined_model_pressure}
\end{figure}

Fig.~\ref{fig:confined_model_density_dist} also probes the chain--end
enrichment effect for confined surfaces. Such an enrichment
has been noticed by Wang and Binder~\cite{Wan91a} in a lattice model.
Shown with the density profile is the relative frequency of end--monomers.
This is computed by defining slices of height $d$ in the $z$-direction
and dividing the number of chain ends by the total number of monomers
within each slice. The resulting relative frequency is fairly independent
of the slice height $d$. We conclude from the figure that there is {\em no}
appreciable chain--end enrichment at confined surfaces. This is in
accordance with the fact that there are no energetic or entropic reasons
for the end--monomers to gather in the vicinity of the plates (see the
discussion for free surfaces below).
The constraints imposed by the plates can favour a high positive virial
contribution to the pressure. Remember that for confined surfaces, the
volume and therefore the mean density of the simulation system is
prescribed. The high pressure in Fig.~\ref{fig:confined_model_pressure} can
be traced back to repelling intermolecular interactions caused by dense
packing.

%%% ---------------------------------------------------------------------------
\section*{Free polymer surfaces}
%%% ---------------------------------------------------------------------------

\begin{figure}
\epsfxsize=14cm
  \begin{center}
    \begin{minipage}{\textwidth}
      \epsffile{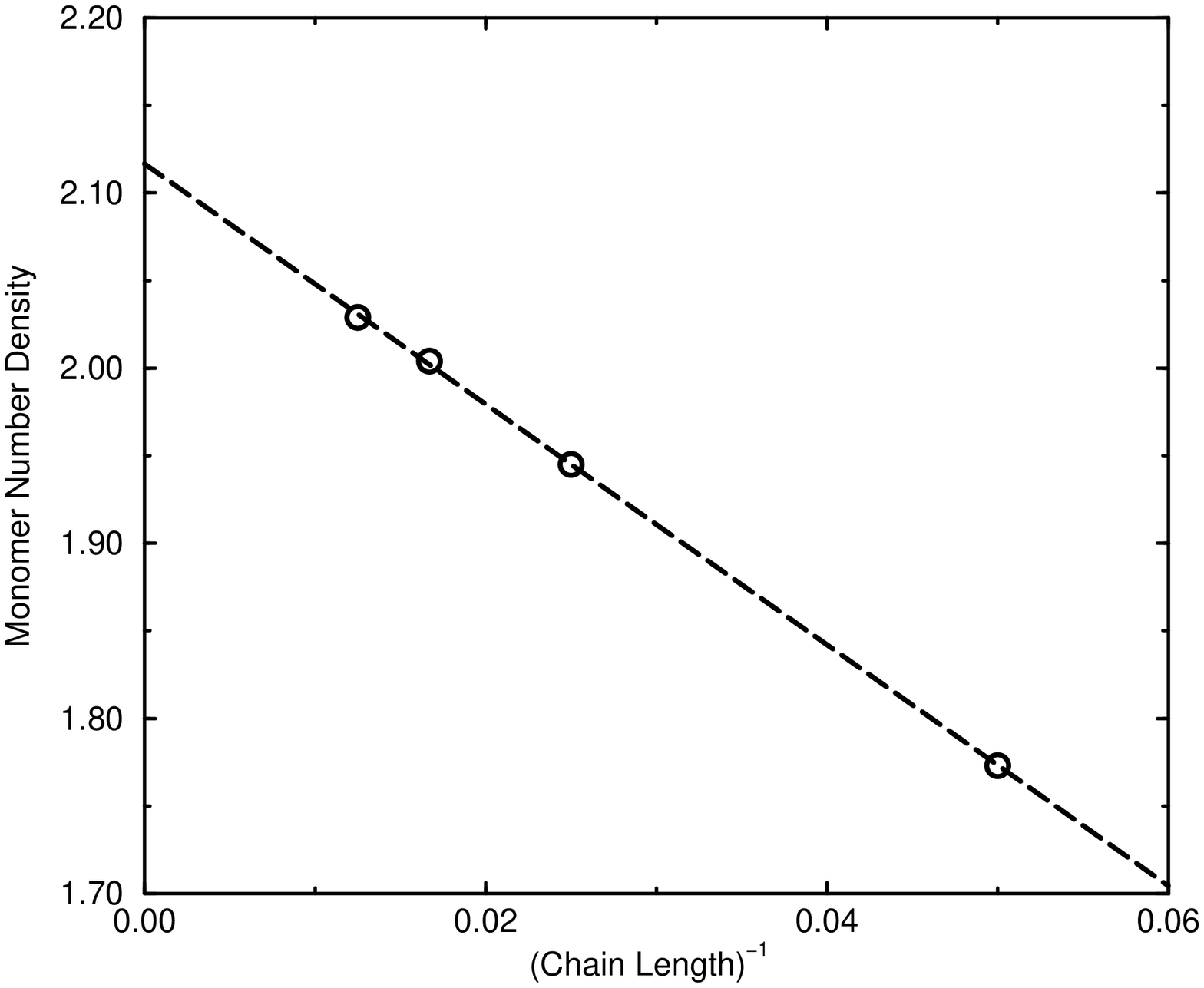}
    \end{minipage}
  \end{center}
\caption{The monomer number density in the bulk (open circles) for
  different chain lengths (20, 40, 60, and 80 monomers) plotted
  vs.~the reciprocal chain length in the {\em free} surface model. The dashed
  line is a linear fit to the data points. Its intercept with the
  $y$--axis yields $\rho^*_{\text{bulk}} ( N \to \infty )
  \approx 2.12$ ($\text{slope} \approx -6.87$).}
\label{fig:free_model_dide}
\end{figure}

To study the effects of a rougher and presumably more realistic surface,
we set up a free system with periodic boundary conditions in $x$-- and
$y$--direction and free boundary conditions in $z$--direction. The polymer
bulk is held together by the attractive part of the inter--molecular
Lennard--Jones interaction. At the beginning, this system can be prepared
like a confined one with two repelling plates in the $z$--direction. This
is mainly done to bring the polymer bulk initially into a regular shape.
By previous experience, the density during the constrained phase is chosen
close to the bulk density of a free system. After a time, the plates are
removed and the system is free to relax into the surrounding vacuum. Another
possibility is to start with a periodic box completely filled with polymer.
This system is run over a relaxation phase which is finished when the chains
have diffused over at least two radii of gyration. After that the periodic
boundary condition in the $z$--direction is removed and the box is rendered
three times as high as before. The polymer film is placed into the middle of
the box and the simulation continues. The density profiles
quickly approach their stationary shape and we made sure that their
bulk or plateau values are independent from the initial density in the
periodic box. Especially for longer chains this gives a good criterion that
the dynamics have approached the equilibrium stage. The bulk densities found
for varying chain lengths are summerized in Fig.~\ref{fig:free_model_dide}.
The reason for the different bulk densities is the ratio of chain--end
to mid--chain monomers. In our model, the effective volume of chain--end
monomers is higher than that of mid--chain monomers, since end--monomers
with only one bond need more room than those with two bonds (the
Lennard--Jones range $\sigma$ is the same for all monomers and longer than
the bond length $l_0$) and also since the end--monomers have a higher mobility
(entropic effect). For different chain lengths, the bulk densities are
interpolated by the formula
\begin{equation}
  \rho_{\text{bulk}} = \frac{m_{\text{chain}}}{V_{\text{chain}}}
       =  \frac{m_{\text{monomer}} N}
          {2 V_{\text{end}} + (N-2) V_{\text{mid}}}
       \approx \frac{m_{\text{monomer}}}{V_{\text{mid}}}
          \left[ 1 - \frac{2 (V_{\text{end}} - V_{\text{mid}})}
          {V_{\text{mid}}} \frac{1}{N} \right] \, .
\end{equation}
Since $V_{\text{end}}-V_{\text{mid}}>0$, there is a $1/N$ chain--end effect
with a negative slope, in accordance with Fig.~\ref{fig:free_model_dide}.
Moreover, note that in the period of time accessible to our simulations,
no chains escape the polymer bulk. Therefore the effective pressure in the
vacuum phase is zero (see also below, Fig.~\ref{fig:free_model_pressure}).

%%% ---------------------------------------------------------------------------
\subsection*{Density profile, chain--end enrichment, and pressure}
%%% ---------------------------------------------------------------------------

\begin{figure}
\epsfxsize=14cm
  \begin{center}
    \begin{minipage}{\textwidth}
      \epsffile{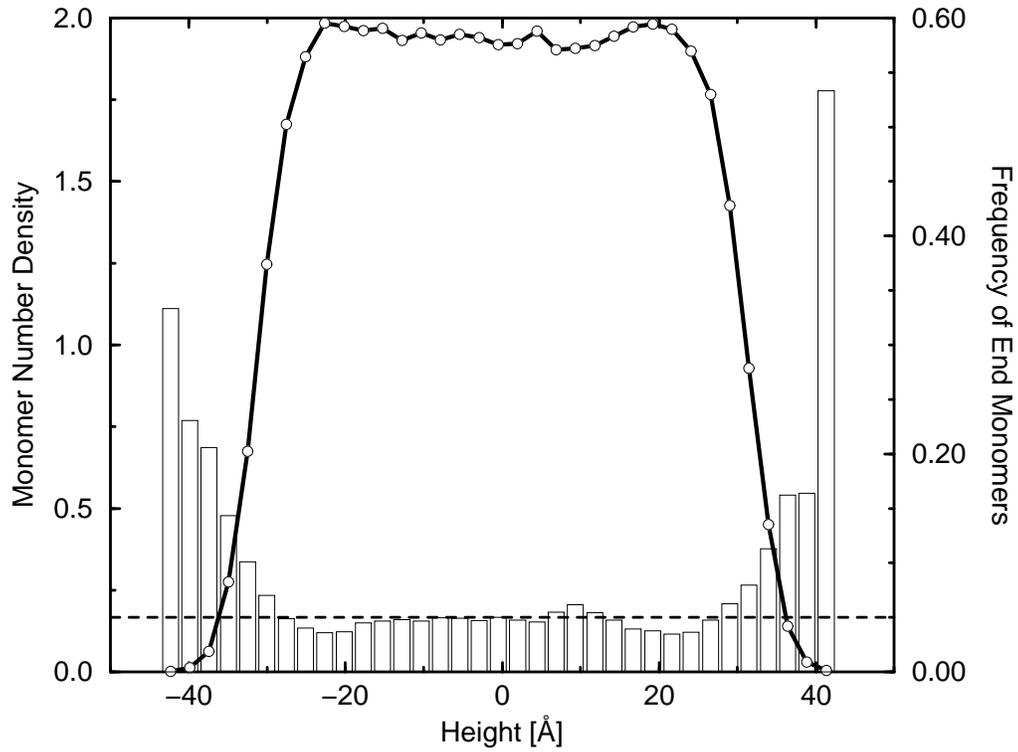}
    \end{minipage}
  \end{center}
\caption{Density distribution and chain--end enrichment for the
  {\em free} surface. Shown is the number density of monomers
  (solid curve, left axis) and the relative frequency of chain--end
  monomers (bar chart, right axis) in slices parallel to the surface.
  The polymer system is monodisperse with a chain length of $N=40$.
  The horizontal dashed line indicates the mean value for chain--ends:
  $\text{2 ends}/\text{40 monomers} = 0.05$.}
\label{fig:free_model_density_dist}
\end{figure}

In Fig.~\ref{fig:free_model_density_dist}, the distribution of the
monomer number density vs.~the height ($z$) direction (the direction normal
to the surfaces) is plotted. As opposed to confined surfaces, no partial 
crystallization can be observed at free surfaces. Also shown in this
figure is the chain--end enrichment. Again in contrast to confined
surfaces, we observe the clear tendency for the chain--ends to
escape from the bulk. Chain--end monomers require a greater static
volume than mid--chain monomers. To reach close packing, they stick
to the surface. This is also favoured by reasons of entropy, since the
end--monomers possess a higher mobility (they are only constrained by
one chemical bond). These arguments, however, do not apply if the surface
is pressed by repelling plates. In Fig.~\ref{fig:free_model_density_dist}
one clearly sees the enrichment effect, but an exact quantitative evaluation
is out of reach, since if we increase the number of bins next to the surface,
the bin width becomes too small to achieve good statistics. The presence of
end--monomers of high mobility and the local brush--like structure caused by
the chain ends will certainly affect the mechanics of surface disturbances
like in an indentation process.

\begin{figure}
\epsfxsize=14cm
  \begin{center}
    \begin{minipage}{\textwidth}
      \epsffile{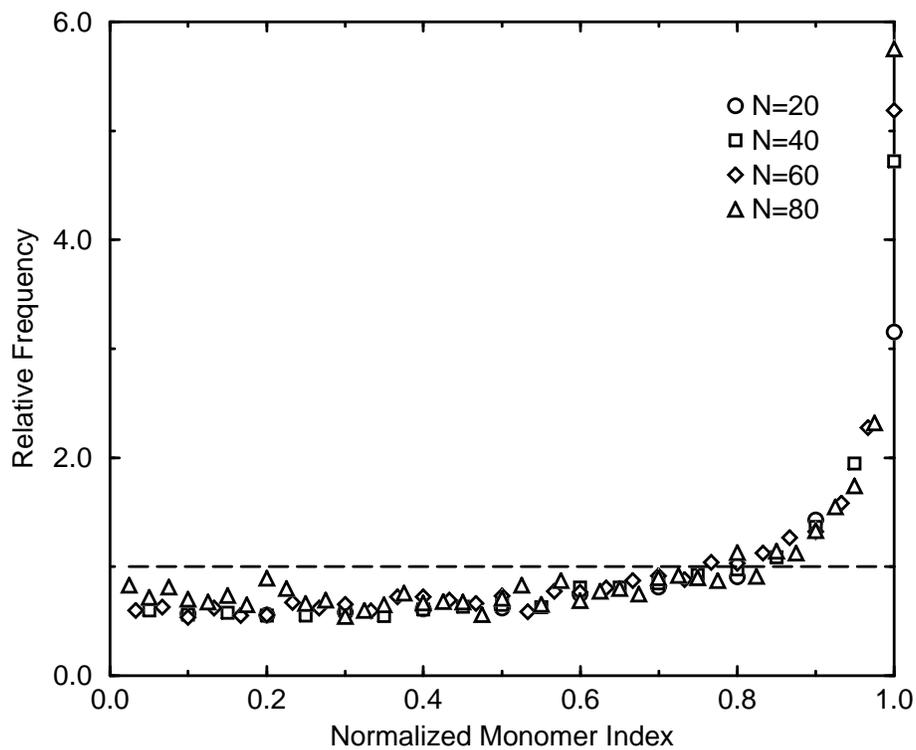}
    \end{minipage}
  \end{center}
\caption{The relative frequency of the normalized monomer index at
  {\em free} surfaces for different chain lengths. If all monomers in
  a chain would appear with the same probability at the surface, the
  relative frequency would be $1$, independent of the normalized index
  (dashed horizontal line). However, higher normalized indices are more
  frequent, a signature of the chain--end enrichment at free surfaces.}
\label{fig:free_model_inde}
\end{figure}

A complementary way to analyze chain--end enrichment is to look at the
distribution of the normalized monomer index $\tilde{\i}$ which is defined as
\begin{equation}
  \tilde{\i} = \frac{| \, i - N/2 \, |}{N/2}
             = \left| \, \frac{2 i}{N} - 1 \, \right| \, .
\end{equation}
Mid--chain monomers get the normalized index $0$, for end--monomers 
it is $1$. Let $n(\tilde{\i})$ be the absolute frequency of the
normalized index $\tilde{\i}$ found at the surface and
$n_{\text{total}} = \sum_{\tilde{\i}} n(\tilde{\i})$ the total number
of all monomers at the surface. These quantities can be determined by
dividing the surface into patches, i.e.~dividing the whole film into
columns. Within each column, we look for the extremal monomers (in a
sense, these monomers then define the surface). The same approach is
chosen below to sample the surface height function. The relative, chain
length independent frequency of $\tilde{\i}$ is then computed according to
\begin{equation}
  \tilde{n} ( \tilde{\i} ) = \frac{n(\tilde{\i})}{n_{\text{total}} / (N/2)}
    = \frac{N n(\tilde{\i})}{2 n_{\text{total}}} \, .
\end{equation}
The result for chain lengths $N=20$, $40$, $60$, and $80$ is shown
in~Fig.~\ref{fig:free_model_inde}. If each monomer in the chain would
appear with the same probability at the surface, the relative frequency
would be $1$, independent of the normalized index (dashed line in the
figure). We observe, however, a significant increase in frequency starting
with a normalized index of around $0.8$. Moreover, the shape of the curves
is very similar for the different chain lengths.

\begin{figure}
\epsfxsize=14cm
  \begin{center}
    \begin{minipage}{\textwidth}
      \epsffile{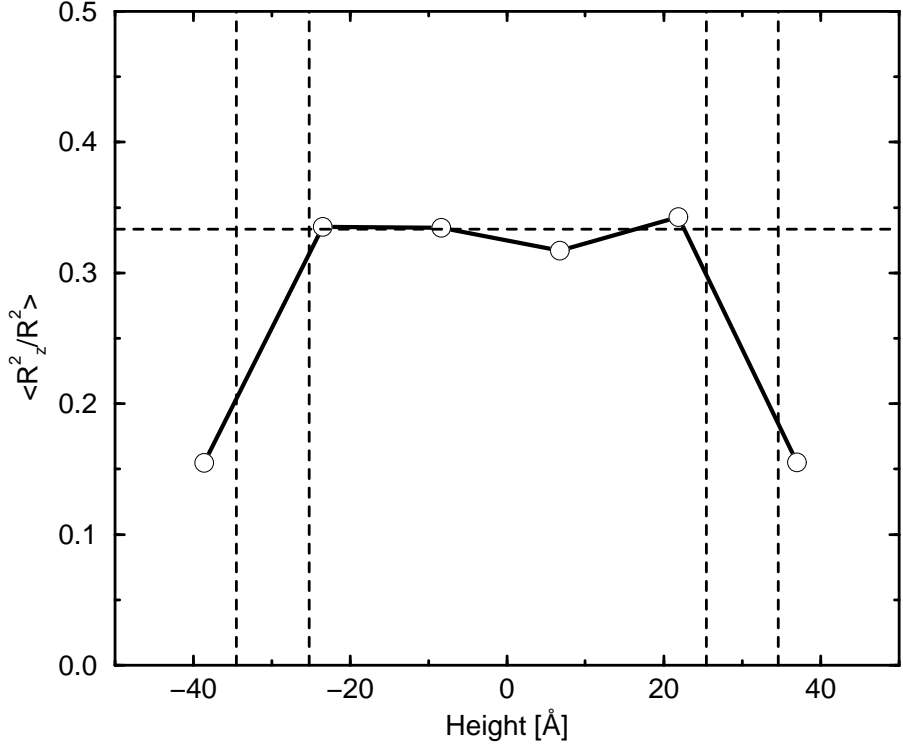}
    \end{minipage}
  \end{center}
\caption{The orientation of the end--to--end vector in the {\em free}
  model. Plotted is the mean square value $\langle c^2_\perp (z) \rangle
  = \langle {R^2_z}/{|\vec{R}|^2} \rangle$
  for the normal--to--surface component after normalization. The mean
  values are computed in slices
  parallel to the surfaces at different height. In the bulk, we expect
  due to equipartition of all directions that $\langle {R^2_z}/{|\vec{R}|^2}
  \rangle_{\text{bulk}} = 1/3$ (horizontal dashed line). The vertical
  lines mark the extension of the surfaces as given by the 10/90--rule
  applied to the number density profile
  in~Fig.~\ref{fig:free_model_density_dist}. The average length of the
  end--to--end vector is found to be 19 {\AA}. This represents a lower
  bound for the bin width in the figure and severely limits the number
  of independent data points.}
\label{fig:free_model_correlation}
\end{figure}

\begin{figure}
\epsfxsize=14cm
  \begin{center}
    \begin{minipage}{\textwidth}
      \epsffile{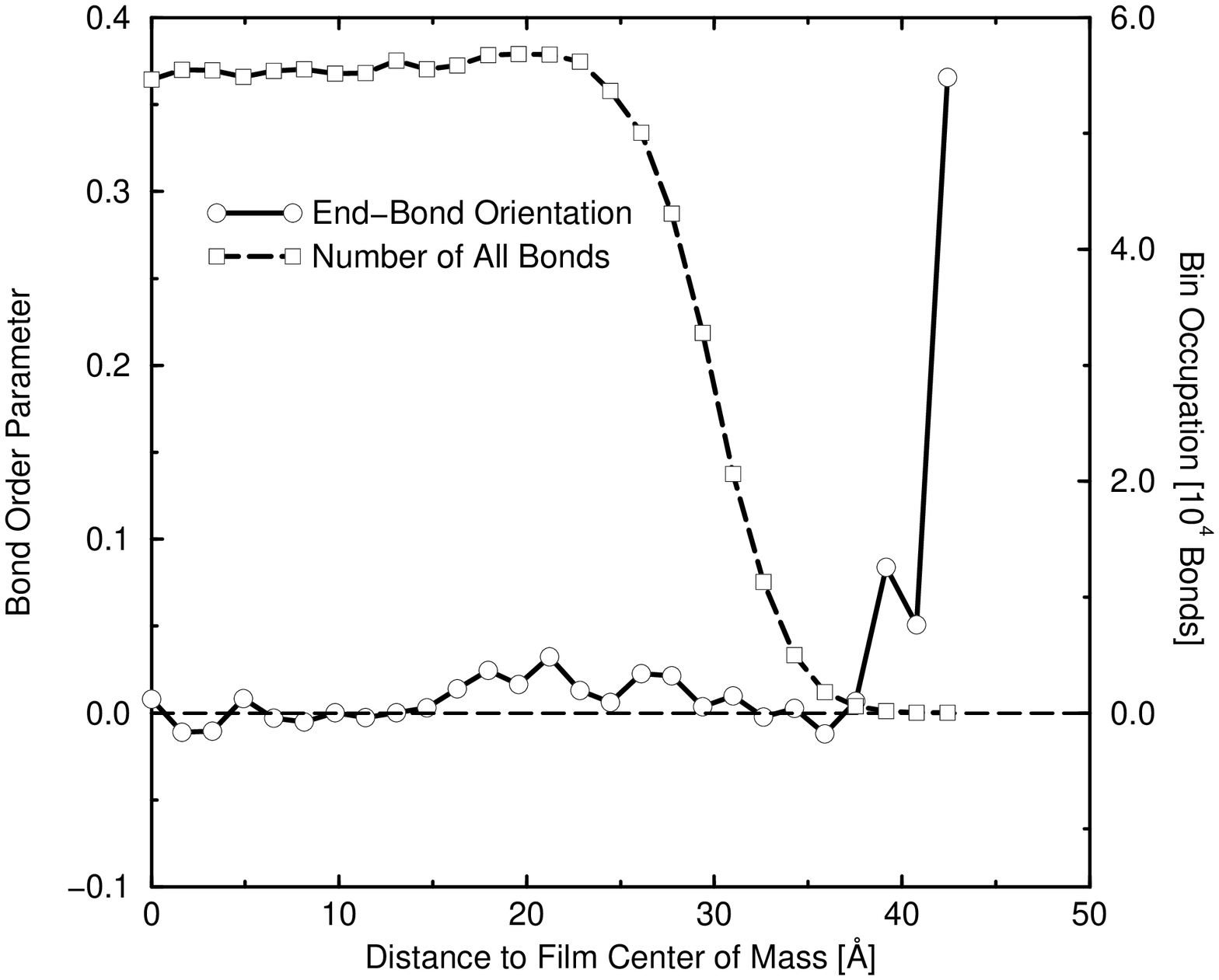}
    \end{minipage}
  \end{center}
\caption{The bond order parameter for the chain--end bonds in the
  {\em free} model. The orientation of the end--bonds is plotted
  vs.~their distance from the system's center of mass  (left axis).
  This involves an average over both surfaces of the film. Good
  statistics for the end--bonds is difficult to obtain. In chains
  of length $N=40$, only 5 percent of the monomers are chain--ends.
  In addition, the number of all bonds becomes very small in the
  surface region, as can be seen in the bin occupation curve (right axis).}
\label{fig:free_model_bondorder}
\end{figure}

If the chain--ends migrate to the surface, one might suspect some
correlation between the orientation of the surface and the chain
end--to--end vectors. This is analyzed in
Fig.~\ref{fig:free_model_correlation}. For a given chain, let
$\vec{R}_{{\text{i}}}$ denote the monomer positions. Then
$\vec{R} = \vec{R}_{\text{N}} - \vec{R}_{\text{1}}$ is the
end--to--end vector. As before, the surfaces are parallel
to the $xy$--plane. We want to look at the normalized projections
of the end--to--end vector normal and parallel to the surface,
\begin{equation}
  c^2_\perp (z) = \frac{R^2_z}{|\vec{R}|^2}
  \qquad \text{and} \qquad
  c^2_\parallel (z) = \frac{R^2_x + R^2_y}{|\vec{R}|^2} \, .
\end{equation}
To place these projection coefficients into bins along the $z$--direction
for statistical analysis, we use the mid--point of the end--to--end vector,
$z = (R_{\text{N}z} + R_{\text{1}z}) / 2$.
Fig.~\ref{fig:free_model_correlation} shows that near the surface, the
end--to--end vectors are on average parallel aligned to it. This indicates
that both end--monomers of a chain are at the same surface. The alignment,
however, does not extend into the plateau region of the density profile. This
is an important result, since for thin films, orientational correlations could
prevent the formation of a bulk zone at all. Furthermore, the average length
of the end--to--end vector in the simulations is 19 {\AA}. Compared to the
film height of about 80 {\AA}, there is no separation of scales.
A related measure for orientation is the bond order parameter, $b(z)$,
defined as
\begin{equation}
  b (z) = \frac{\langle b_z^2 \rangle - \frac{1}{2} (\langle b_x^2 \rangle
  + \langle b_y^2 \rangle)}{\langle |\vec{b}|^2 \rangle} \, ,
\end{equation}
where $\vec{b}$ with components $b_x$, $b_y$, and $b_z$ denotes an arbitrary
bond vector connecting two monomers. In Fig.~\ref{fig:free_model_bondorder},
the average is over chain--end bonds only. The mean values are computed
from a number of statistically independent configurations in bins (slices)
of finite  width parallel to the film surfaces ($xy$--plane). The bond order
parameter is plotted vs.~the distance of the bins from the center of mass.
In the bulk, we expect isotropic behaviour, yielding $b = 0$. We want to test,
whether the chain--end bonds in the vicinity of the surface are oriented
perpendicular to it, which means $b_z^2 > b_{x/y}^2$ resulting in $b > 0$.
Indeed, from Fig.~\ref{fig:free_model_bondorder} we infer that this is
the case. However, sufficient statistical data for the orientation of the
end--bonds at the surface are difficult to obtain. The reason is that we
sample in a region in space where the density is already very low.

\begin{figure}
\epsfxsize=14cm
  \begin{center}
    \begin{minipage}{\textwidth}
      \epsffile{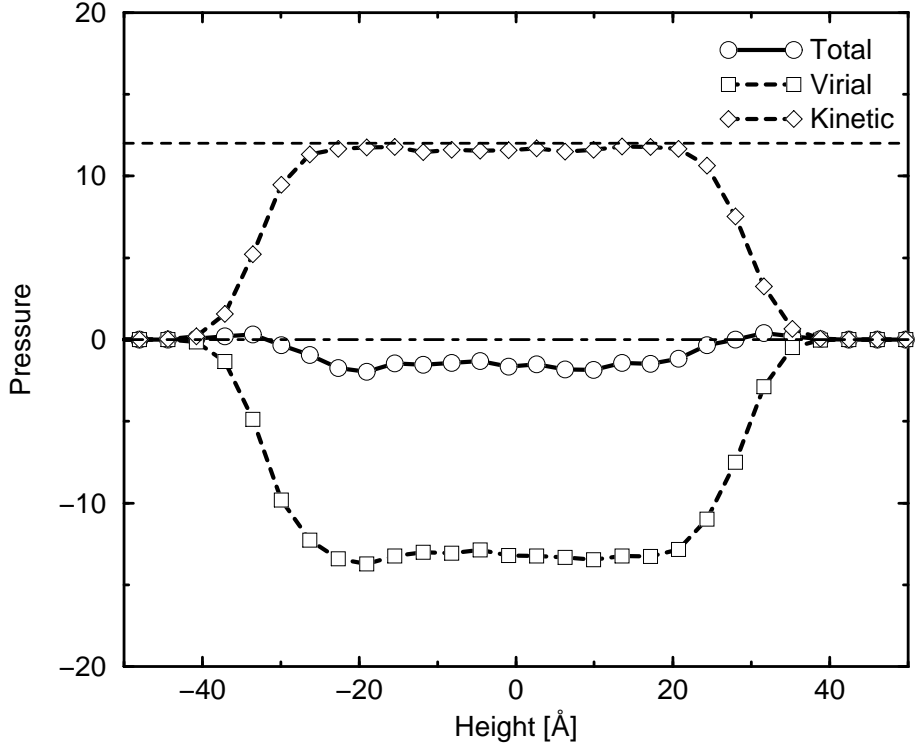}
    \end{minipage}
  \end{center}
\caption{The pressure distribution vs.~the height direction for the
  {\em free} surface model (pressure in reduced units). The total
  pressure is the sum of the virial and the kinetic contribution.
  The dashed horizontal line indicates the expected value for the
  kinetic (ideal) part: $p^{*}_{\text{ideal}} = \rho^* T^* \approx
  2.0 \cdot 6.0 = 12.0$. The dot--dashed line is the zero pressure line.}
\label{fig:free_model_pressure}
\end{figure}

Fig.~\ref{fig:free_model_pressure} shows the pressure distribution.
In effect, a free surface is the interface to a vacuum. This means
that the external pressure imposed on the free polymer film is zero.
Kinetic and virial contributions should therefore compensate each
other. In the figure, the total pressure is slightly less than zero
and the system will still contract somewhat during a final relaxation
phase that has not been reached in the simulation.

%%% ---------------------------------------------------------------------------
\subsection*{Surface profiles and surface width}
%%% ---------------------------------------------------------------------------

In this section, we are concerned with possible analytical fits for surface
profiles and the systematic determination of surface location and width.
A first idea is to start from the (number or mass) density profile in
Fig.~\ref{fig:free_model_density_dist} and to look for the points
$z_{10}$ and $z_{90}$ where 10 and 90 percent of the bulk density are
reached. The surface width is the distance between these two points,
$\xi = | \, z_{10} - z_{90} \, |$ and as the surface base point we choose
the the midpoint, $z_0 = (z_{10} + z_{90})/2$. In what follows, this
method will be called the ``10/90--rule''.
A second possibility is to fit the density profile with some
suitably parameterized analytic function and to use these parameters
to actually characterize the surface. This allows to compare profiles which
have been fitted using the same family of functions. An appropriate family
is based on the hyperbolic tangent,
\begin{equation}
  \rho (z) = \frac{1}{2} \rho_{\text{bulk}} \left( 1 - \tanh
    \frac{z-z_{\text{t}0}}{\xi_{\text{t}}} \right) \, ,
\end{equation}
which is essentially the Fermi function, since
$(1/2) (1 - \tanh x) = (1 + \exp 2x)^{-1}$. The 10/90--rule applied to
a $\tanh$--fit yields the relations $z_{\text{t}0} = z_0$ and
$\xi_{\text{t}} = \xi / \ln (0.9/0.1) \approx 0.46 \, \xi$.
Such a fit is used in Fig.~\ref{fig:free_model_coor}.
Another possibility is to use the error function (the probability
distribution for the Gaussian normal probability density) for the fit,
\begin{equation}
  \rho(z) = \rho_{\text{bulk}} \int_{- \infty}^z f(\zeta) \, d \zeta 
  \qquad \text{with} \qquad
  f(z) = \frac{1}{\sqrt{2 \pi} \sigma} \exp \left( -
    \frac{(z-z_{\text{g}0})^2} {2 \sigma^2} \right) \, .
\end{equation}
The connection to the 10/90--rule is given by $z_{\text{g}0} = z_0$ and
$\sigma \approx  2.56 \, \xi$. Note that these functions also appear in the
following paragraph when characterizing the surface by use of a ``height
function'', but then they enter in a purely statistical context.

\begin{figure}
\epsfxsize=14cm
  \begin{center}
    \begin{minipage}{\textwidth}
      \epsffile{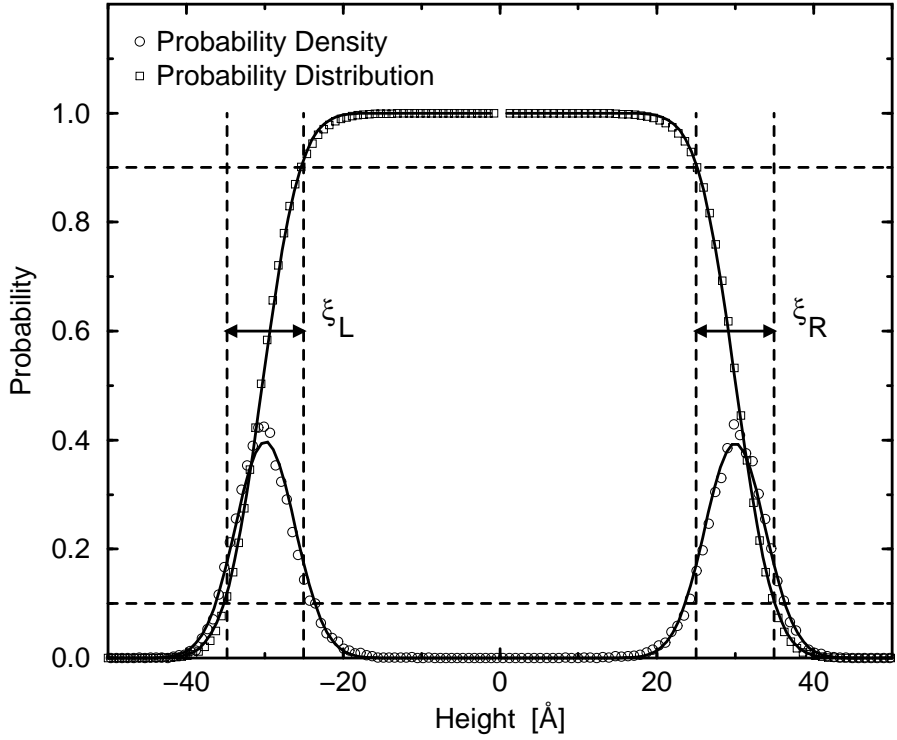}
    \end{minipage}
  \end{center}
\caption{The {\em free} surface profile as obtained by the statistical
  analysis of the height function. The circles mark the probability to
  find the highest (or lowest, respectively) monomer of a column at the
  given height. This probability density comes from the simulation and
  has been integrated numerically to determine the associated probability
  distribution function which is marked by the squares. The probability
  distribution represents the ``surface profile'' in the height function
  approach. The solid curves are analytical fits to the data. For the
  probability, Gaussian normal densities with $\sigma_{\text{L/R}} = 2.56
  \, \xi_{\text{L/R}}$ are used. The surface profile is a properly located
  and scaled error function. The surface widths ($\xi_{\text{L}}$ and
  $\xi_{\text{R}}$) and boundaries (dashed lines) as given by the 10/90--rule
  are also indicated.}
\label{fig:free_model_gaus}
\end{figure}

Because of their steepness, surface characterization based solely on the
flanks of the density profiles is problematic. The profiles must
be determined by sampling monomers. If we put more bins onto the flanks,
the bin width soon becomes too small for sufficient statistics. The following
method offers a complementary approach. We look onto the upper surface from
above and divide it -- in the simplest case by use of a regular grid -- into
a number of patches. Within each patch, (which represents a column in the
three--dimensional film) we identify the monomer with the highest
$z$--coordinate. In this way we obtain a mapping from the patches to a set of
height values. We call this mapping the ``height function''. For the given
set of height values, we determine mean value and standard deviation and
compute from them the location and the width of the surface, respectively.
To get the result for the lower surface, we have to pick the lowest monomer
within each column. The method is related to the surface contour one obtains
using a scanning force microscope. A resolution of 1 or 2 Lennard--Jones
$\sigma$ for the resolution (distance between grid lines) is found to be
appropriate. When we plot the relative frequency of height values found, we
obtain a probability density as in Fig.~\ref{fig:free_model_gaus} which can
be fitted with a Gaussian normal density. Then the corresponding distribution
function is determined which represents some average surface profile.

\begin{figure}
\epsfxsize=14cm
  \begin{center}
    \begin{minipage}{\textwidth}
      \epsffile{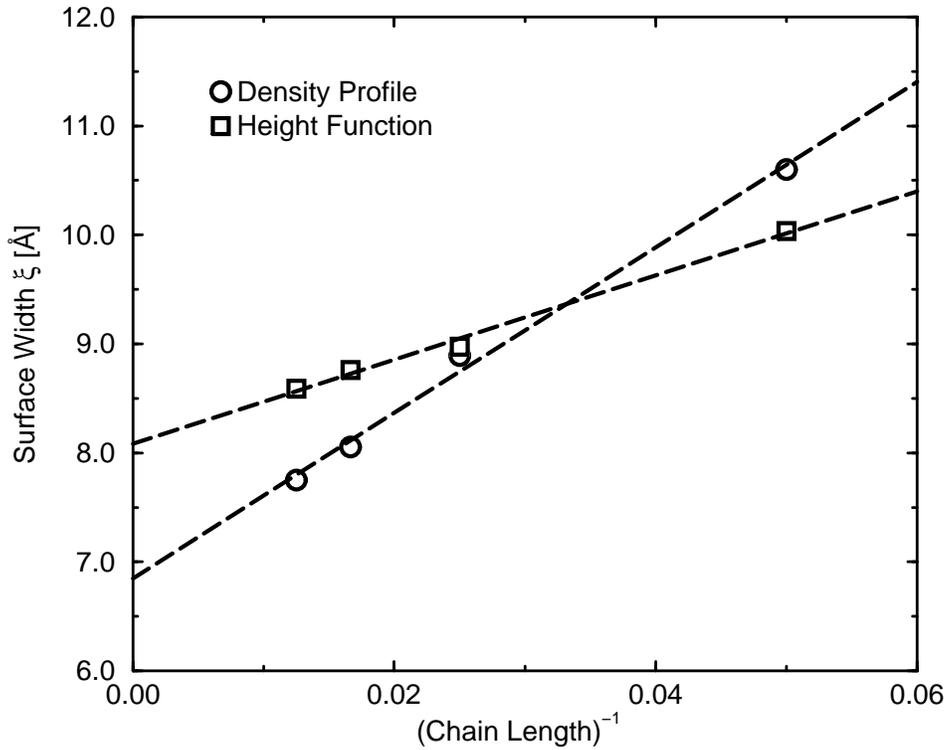}
    \end{minipage}
  \end{center}
\caption{The surface width $\xi$ vs.~the reciprocal chain length ($N = 20$,
  $40$, $60$, and $80$) in the {\em free} model. Slightly different values
  for $\xi$ are obtained by applying the 10/90--rule either to number density
  profiles as in Fig.~\ref{fig:free_model_density_dist} (circles) or to
  probability distribution profiles based on the height function as in
  Fig.~\ref{fig:free_model_gaus} (squares). The plot suggests a
  scaling $\xi \sim N^{-1}$. The asymptotic values resulting from linear
  fits (dashed lines) are $\xi \, (N \to \infty) = 6.8 \, \text{{\AA}}$
  for the density profile and $\xi \, (N \to \infty) = 8.1 \, \text{{\AA}}$
  for the height function method.}
\label{fig:free_model_width}
\end{figure}

The relation between the height function profile and the monomer number
density profile can be derived in the special case that the monomer
density assumes its bulk value instantly below the surface. In this
case, the finite slopes in the average density profiles stem only from
the fact that the surface is rough. The surface is defined by the height
function $h(x,y)$ over a domain in the $xy$--plane with base area~$A$.
For the height coordinate~$z$ we obtain the probability density
\begin{equation}
  p_{\text{hf}} (z) = \frac{1}{A} \int \! \! \int dx \, dy \,
    \delta (  h(x,y) - z  ) \, .
\end{equation}
The associated distribution function is
($\Theta \, (\,{z}\,)$ is the unit step function)
\begin{equation}
  P_{\text{hf}} (z) = \int_{- \infty}^{z}
       p_{\text{hf}} ( \zeta) \, d \zeta  
       = \frac{1}{A} \int \! \! \int dx \, dy \, \Theta (  h(x,y) - z  ) \, .
\end{equation}
On the other hand, the mean density at height $z$ is computed by
``counting all columns filled with matter'',
\begin{equation}
  \rho(z) = \frac{\rho_{\text{bulk}}}{A} \int \! \! \int dx \, dy \,
            \Theta (  h(x,y) - z  ) \, .
\end{equation}
Under the condition mentioned in the beginning, one therefore finds
\begin{equation}
  \rho(z) = \rho_{\text{bulk}} P_{\text{hf}} (z) \, .
\end{equation}
Of course, this is what one would have guessed knowing that the density
reaches its constant bulk value instantly below a surface with zero local
thickness.  In Fig.~\ref{fig:free_model_width} the thickness measures
obtained from monomer density profiles and height function statistics are
compared to each other for different polymer chain lengths. The fit
$\xi \sim N^{-1}$ was successful. We find that for longer chains the height
function method, based on the statistics over outermost monomers, certainly
overestimates the surface width.

%%% ---------------------------------------------------------------------------
\subsection*{Coordination number}
%%% ---------------------------------------------------------------------------

\begin{figure}
\epsfxsize=14cm
  \begin{center}
    \begin{minipage}{\textwidth}
      \epsffile{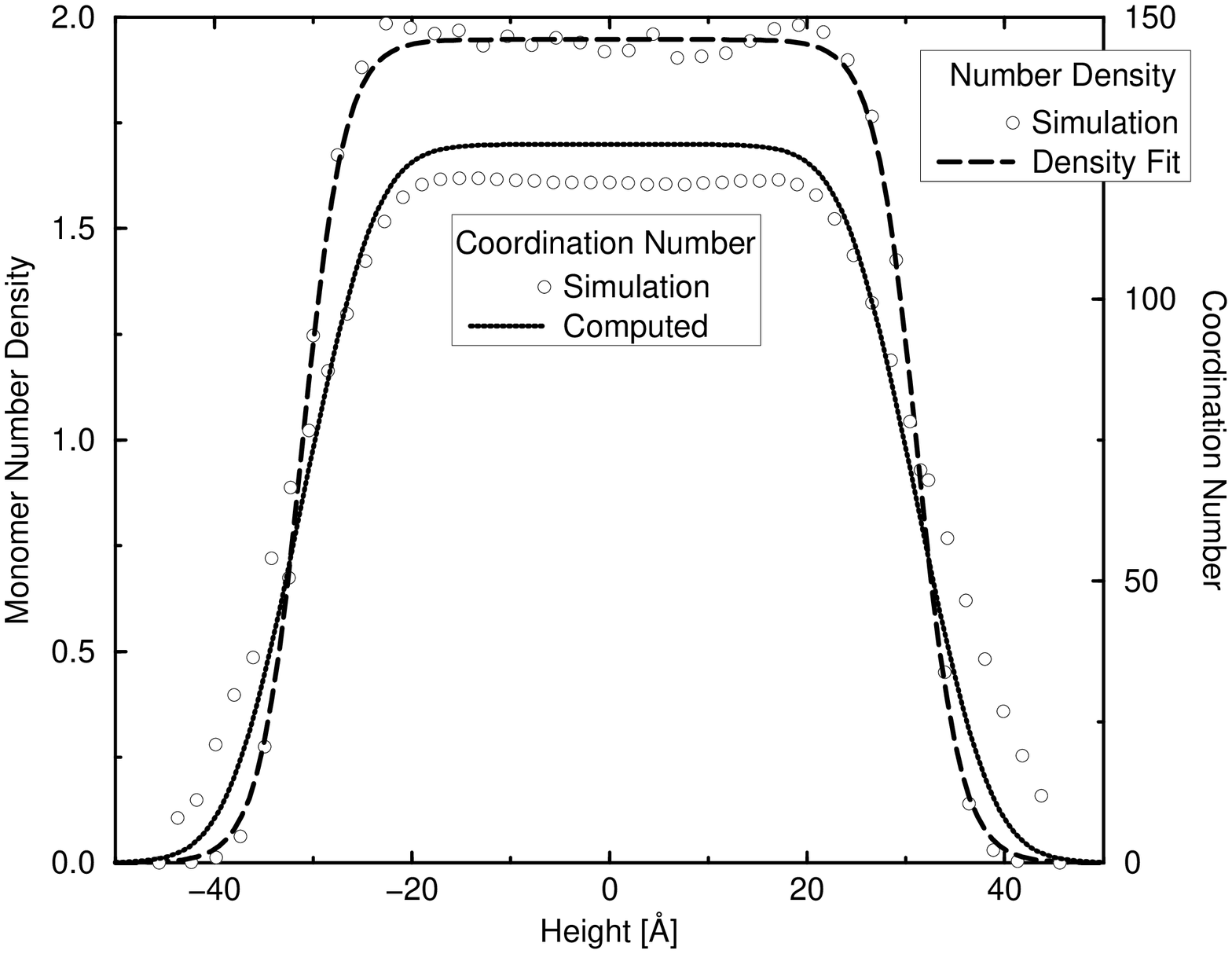}
    \end{minipage}
  \end{center}
\caption{Comparison between density profile and coordination number for
  the {\em free} surface model. The simulated density curve is the same
  as in~Fig.~\ref{fig:free_model_density_dist}. For the density fit, a
  combination of two hyperbolic tangent functions is used. This fit allows
  a theoretical estimate for the coordination number which is given
  in the figure, too. In the worst case, the estimate must be determined
  by a numerical integration, whereas the determination of the coordination
  numbers from the simulation is more time consumptive. Over all sample
  configurations, a double loop must be run, since for each monomer we
  have to count all its interaction partners. Note that the measured
  coordination number curve is even broader than the theoretical one.}
\label{fig:free_model_coor}
\end{figure}

In connection with amorphous materials, the coordination number
of a given particle is the number of all its partners involved in
van--der--Waals (Lennard--Jones) interactions. From the simulations,
this number can simply be determined by counting all the monomers in the
interaction range (a sphere of radius $2.5$ Lennard--Jones $\sigma$) of
a given monomer (monomers in the same chain do count except they take part 
in bond length, bond angle, or torsion interactions). Again we assume
homogeneity in the lateral ($x$ and $y$) directions, associate with
every coordination number the coordinates of its monomer, and set up the
distribution of the average coordination number vs.~the height ($z$)
direction. The result is shown in Fig.~\ref{fig:free_model_coor}.
A theoretical estimate for the coordination number distribution $c(z)$
must be based on the monomer number density profile $\rho(z)$.
Geometric relations then yield the convolution type integral
\begin{equation}
  c(z) = {\displaystyle \int_{{z} - { \sigma}}^{{z}
    + { \sigma}}} { \rho}(\,{\zeta}\,)\,{ \pi}\, \left( \! \,{ \sigma}^{
      2} - (\,{r} - {\zeta}\,)^{2}\, \!  \right) \,{d}{\zeta} \, .
\end{equation}
For the monomer density profile, $\rho (z)$, we use the ansatz
\begin{equation}
  \rho(z) = \rho_{\text{bulk}} \left[ \Theta \, (\, - {z}\,) \,
    \left( \! \,1 + {\rm tanh} \left( \! \,{ \displaystyle \frac {z
    + z_{ \text{tL}}} {\xi_{\text{tL}}}} \, \!  \right) \, \!  \right)  
    + \Theta \, (\,{z}\,)\, \left( \! \,1 - {\rm 
    tanh} \left( \! \,{\displaystyle \frac {z - z_{\text{tR}}}
    { \xi_{\text{tL}}}} \, \! 
    \right) \, \!  \right) \right]
\end{equation}
with the parameters $\sigma = 9.8 {\,\text{\AA}}$,
$\rho_{\text{bulk}} = 1.947$,
$z_{ \text{tL}} = 31.16 {\,\text{\AA}}$,
$z_{ \text{tR}} = 31.16 {\,\text{\AA}}$,
$\xi_{\text{tL}} = 4.32 {\,\text{\AA}}$,
and $\xi_{\text{tR}} = 4.32 {\,\text{\AA}}$.
This results in the computed curve for the coordination number in
Fig.~\ref{fig:free_model_coor}. The discrepancy in the plateau value
derives from the fact that in the estimate we implicitly also count
the neighbor monomers in the chain (which should be neglected, see
above). As expected, the flanks of the coordination number distribution
are broader than the flanks of the density profile. This means that
in the interaction picture, the surface region is further extended than
according to the local density criterion.

%%% ---------------------------------------------------------------------------
\section*{Conclusion and outlook}
%%% ---------------------------------------------------------------------------

We have presented two methods to prepare amorphous polymer surfaces in
molecular dynamics simulations with united atom models for the polymer
chains. The ideas involved can easily be transfered to both more detailed
models (which, as one extreme, explicitly treat each atom in a monomer,
or, like the ellipsoidal model~\cite{Zim95a} for BPA--PC, extend the
united atom model by considering the geometric shape of a
monomer) and more coarse--grained polymer models. The use of classical
molecular dynamics algorithms is not mandatory. Monte--Carlo and Hybrid
Monte--Carlo techniques can be employed as well and the inclusion of
stochastic Langevin forces (Brownian motion) is straightforward (and,
in fact, implemented in our simulation program).
The present work is also a further step in extending the versatility of
{\em The Materials Explorer}, a continuously growing, integrated software
tool for the simulation, visualization, and quantitative analysis of
engineering materials~\cite{Hee95a}.
The work presented here will allow us to analyze our simulations on
nano--mechanical surface indentation processes~\cite{Lin96a} in which
we prepare a surface using one of the methods discussed above. 
After that, the impact of the indentation tool causes local disturbations
and the deviation of the certain observables from their equilibrium
values gives us information about the surface resistance and friction
properties. Related to the indentation processes are simulations on
thin film bending~\cite{Pae96a}. A tool, which is more carefully driven 
as in the indentation experiment, causes non--local, large--scale
deformations in the polymer film. Again various observables
(density distribution, orientation of the chain molecules) deviate
from their equilibrium values and have to be analyzed.

%%% ---------------------------------------------------------------------------
\section*{Acknowledgments}
%%% ---------------------------------------------------------------------------

The authors gratefully acknowledge the support from the Bundesministerium
f\"ur Bildung und Forschung (BMBF) in the framework of the project
``Computer Simulation Komplexer Materialien'' under grant No.~03N8008D.
Part of this work was funded by a Stipendium of the Graduiertenkolleg
``Modellierung und Wissenschaftliches Rechnen in Mathematik und
Naturwissenschaften'' at the IWR Heidelberg. We thank Grant~D.~Smith for
valuable discussions.

%%%%%%%%%%%%%%%%%%
%%% References %%%
%%%%%%%%%%%%%%%%%%

%%%%%%%%%%%%%%%
%%% Figures %%%
%%%%%%%%%%%%%%%

%%%%%%%%%%%%%%
%%% Tables %%%
%%%%%%%%%%%%%%

\end{document}